\documentclass[aip, apl, superscriptaddress, reprint]{revtex4-1}
\usepackage{graphicx} 
\usepackage{bm}
\usepackage{natbib}
\usepackage{verbatim}
\usepackage{epstopdf}
\usepackage{hyperref}

\begin{document}
	\preprint{AIP/123-QED}
	
	\title{Derivation of potential profile of a dynamic quantum dot}
	
	\author{N. Johnson}
	%\email[]{johnson.nathan.sd@hco.ntt.co.jp}
	\affiliation{NTT Basic Research Laboratories, NTT Corporation, 3-1 Morinosato Wakamiya, Atsugi, Kanagawa, 243-0198, Japan}
	\author{G. Yamahata}
	\affiliation{NTT Basic Research Laboratories, NTT Corporation, 3-1 Morinosato Wakamiya, Atsugi, Kanagawa, 243-0198, Japan}
	\author{A. Fujiwara}
	\email[]{akira.fujiwara.kd@hco.ntt.co.jp}
	\affiliation{NTT Basic Research Laboratories, NTT Corporation, 3-1 Morinosato Wakamiya, Atsugi, Kanagawa, 243-0198, Japan}

	\date{\today}
	
	\begin{abstract}
	We report a method to derive the potential barrier profile shape in a dynamic quantum dot and show the loading statistics, and hence accuracy of electron transfer, depend significantly on the shape of the barrier. This method takes a further step towards tunable barrier shapes, which would greatly increase the accuracy of single electron sources, allowing the single electron current to be useful for quantum sensing, quantum information and metrology. We apply our method to the case of a tunable-barrier single-electron pump, an exemplary device that shows promise as a source of hot single electron wavepackets.
	\begin{comment}
		
	By measuring the temperature dependence of the accuracy of electron transfer in a tunable-barrier single-electron pump, we deduce the cross-over temperature $T_{0}$, and by extension derive the best possible accuracy for our device. This previously-unemployed technique can be used to rapidly map the minimal error of single electron sources, without recourse to technically-challenging direct error evaluation. Further, we show how our analysis allows us to determine other important accuracy limiting parameters that must be accounted for in the design process, such as the cross-coupling terms. Finally, we combine this information to evaluate the charging energy $E_{c}$, which has been previously difficult to evaluate in the dynamic quantum dots used here.
	\end{comment}
	\end{abstract}
\maketitle

Dynamic quantum dots (dQD) have been demonstrated to be able to capture single electrons with very high accuracy \cite{Blumenthal,GiblinReview, Giblin, Bae, Yamahata, Zhao,Rossi,Stein}.
Proposals for their application include roles in sensors \cite{Johnson,Yamahata5, joanna, Bocquillon2}, metrological current devices \cite{Zimmerman, Giblin2, Scherer, Kaneko}, quantum information schemes \cite{Bauerle} and electron quantum optics \cite{Jon, JonNatNano}.
In each of these cases, the dQD provides a high-accuracy, high-fidelity source of single electron wavepackets.
One promising architecture of a dQD system that can be embedded in the above mentioned proposals is the tunable-barrier single-electron pump. 
This is a system in which a barrier with a time varying potential creates a dQD, isolates a single electron within the dQD, and ejects the resident electron to the drain, within each cycle of the time varying potential.
The forward-ejected electrons create a dc current. 
This out-of-equilibrium architecture is expected to be able to fulfil the criterion of becoming a realisable metrological standard at an accuracy level of $10^{8}$ \cite{GiblinReview} and become a leading class of single electron source.
To date, however, studies of its detailed working, which would lead to better development of this class of devices, have been empirically led. 
In this work we begin to remedy this by deriving the potential profile of the time-varying barrier for the first time, which determines the loading process, and hence performance of a pump or dQD.	
This can lead to better informed lithographic designs and operation, making the dQD system far more applicable to a wide class of single-electron based technologies.

In a dynamic quantum dot system, electrons are captured in a quantum dot (QD) by crossing a time-dependent potential barrier.
%These electrons are later ejected from the dQD, creating a current.
In the tunable-barrier single-electron pump, the dQD is emptied by ejecting electrons over a fixed static barrier.
Under the correct operating parameters, this ejection process can be made to occur with unit probability, making the accuracy of the resultant current dependent only on the loading process \cite{Slava,Fujiwarabook,Kaestner2}.
The accuracy of a dQD can be defined as the fidelity of capturing a fixed given number of electrons.
During the loading process, the dQD population is sensitive to tunnelling across the dynamic barrier and thermal fluctuations.
Hence, understanding the dynamic barrier potential profile at the time of loading (see Fig.~1(b)) is crucial for creating the highest accuracy devices, and establishing a high fidelity single electron source.
Further, we note this characteristic cannot be derived from the single electron transistor (SET) behaviour of a device, because of the time-dependent energy scale.
In an SET, the main parameters determining the threshold for conduction are the temperature and the addition energy. 
In a dynamical system, we counter that the role of these parameters is somewhat different because of the interplay between the thermal excitation and barrier tunnelling.

In this letter we derive the potential profile of the dQD of a tunable-barrier single-electron pump by measuring the temperature dependence of the transfer accuracy.
We propose this as a general method that can be used to evaluate design critical parameters of a dQD, such as the barrier shape, charging energy, and cross-coupling between the dynamic barrier and quantum dot.
All of these parameters will affect the absolute accuracy of the dQD system, and the fidelity of the system to act as a source of tunable size wavepackets \cite{Jon,JonNatNano,Yamahata5,JohnsonPRL,Yamahata3}.
We note that whilst the model accuracy of an individual device can be evaluated straightforwardly \cite{Kaestner2,Slava,GiblinReview}, this value provides no information on how to improve such accuracy or its limiting parameters. 
Here we provide a new dimension of measurement, which we can envisage can become a central technique in leading future device design.

Figure~1(a) shows an image of a device similar to the one studied in this work. 
A Si-nanowire of width 10~nm is spanned by two polycrystalline-Si gates $\rm{G_{1}}$ and $\rm{G_{2}}$, which have length 40~nm and separation 100~nm. 
They are separated from the nanowire by a thermally grown 30~nm oxide layer.
When voltages $V_{ent}$ and $V_{exit}$ are applied to each gate $\rm{G_{1}}$ and $\rm{G_{2}}$ respectively, potential barriers are formed in the channel underneath the gate.
A polycrystalline-Si upper gate covers the complete area in the image, to aid conduction.
Because our device works predominantly using electric field confinement, we suggest our methodology and results discussed here apply equally to dQD constructed in other media \cite{Kaestner2, GiblinReview}. 

\begin{figure}
%\subfigure
	\includegraphics[width=\linewidth, scale=1]{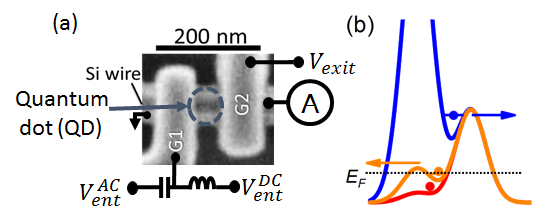}
%\subfigure
	\includegraphics[width=\linewidth, scale=1]{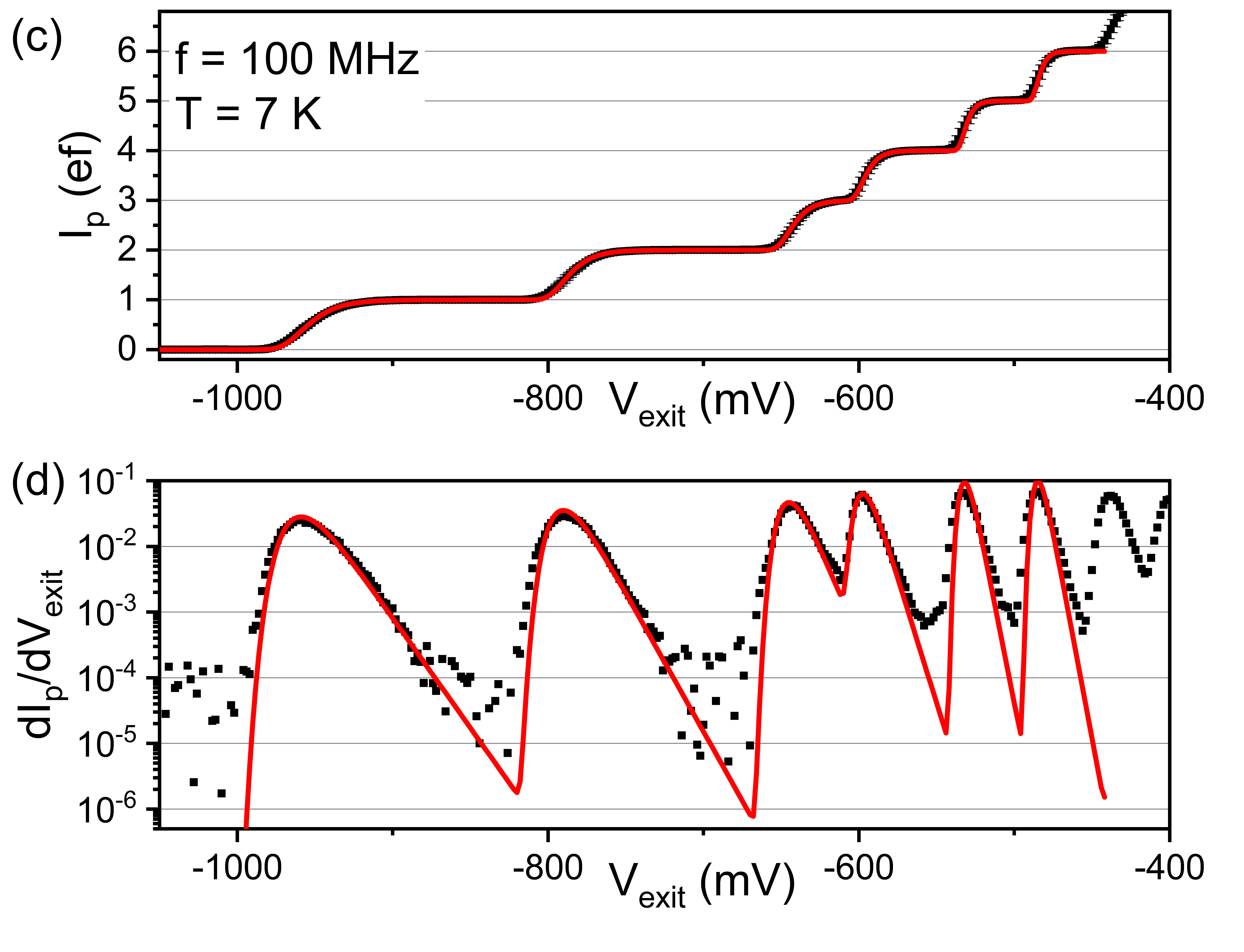}
\caption{
	(a) A silicon tunable-barrier single-electron pump equivalent to the one used in this work. Polycrystalline-Si gates $\rm{G_{1}}$ and $\rm{G_{2}}$ span a Si nanowire, and can be used to form potential barriers, forming a QD in between them in the case of single electron pump operation. 
	(b) Principle of operation of the electron pump. $\rm{G_{1}}$ is driven by the ac voltage $V_{ent}^{AC}$ whilst $\rm{G_{2}}$ is held static by $V_{exit}$. 
	Potential profile defining the QD for three stages in the cycle of $V_{ent}^{AC}$. Red - initial population, orange - escape, blue - ejection. The red and orange stages together determine the loading profile.
	(c) Electron pump transfer current $I_{p} = nef$, with $n=0$ to $6$ shown here (black), and a decay cascade model fit (red).
	(d) Derivative $dI_{p}/dV_{exit}$, with the corresponding derivative of the decay cascade fit in red. 
	}
	\label{Fig.1}
		\end{figure}

%\begin{figure}
%	\includegraphics[width=\linewidth, scale=0.75]{fig2.png}
%	\caption{
%		(a) Potential profile of the tunable-barrier single electron pump at three snapshots in the pumping cycle: capture (red), isolation  (orange), and ejection (blue).
%		(b) Example dc output of the pump $I_{p} = nef$ shows step-like plateaus determined by the charging energy (black), and a fit to the decay cascade model (red) for $n = 1$ shows good agreement.
%		}
%	\label{Fig.2}
%\end{figure}

%Now we consider the case of the single-electron pump.
%In this case, we add an additional sinusoidal ac waveform to $G_{1}$ such that now $V_{G1} = V_{G1}^{AC} + V_{G1}^{DC}$.
$V_{ent}$ is composed of a sinusoidal ac waveform with dc offset $V_{ent} = V_{ent}^{AC} + V_{ent}^{DC}$, whilst $V_{exit}$ is dc only.
The potential profile created by the two gates along the nanowire is sketched in Fig.~1(b).
We separate the period of $V_{ent}^{AC}$ into three key stages of the pump cycle.
In red, the dQD energy is below the Fermi energy $E_{F}$ and electrons can populate the dQD.
As the potential rises (orange), escape to the source occurs to leave a dQD with metastable occupation $n$. 
It is at this stage an error, defined as the occupation of any value of $n$ other than the most probable (including $n = 0$) is most likely to occur.
The loading profile is determined by these two stages.
Finally, this population is frozen in as the dQD is isolated, and the potential barriers are large.
Towards the maximum amplitude of $V_{ent}^{AC}$, the dQD is emptied (blue) and all resident electrons are ejected to the drain.
This creates a dc current $I_{p} = nef$, with $e$ the electronic charge and $f$ the repetition frequency \cite{Fujiwara2}.
By measuring $I_{p}$ and comparing it to the expected value for perfect transfer, we measure the loss of electrons during the escape to source phase, and hence evaluate the error.
We measure our device at $f = 100$~MHz, understood to be sufficiently low to allow the loading process to occur adiabatically \cite{inprep, Kataoka2}.
We use current to voltage conversion of $10^{10}$~V/A using a Femto DDPCA-300 amplifier connected to the drain lead and $I_{p}$ is measured by an HP3458A voltmeter using 10~NPLC averaging to minimise noise.
The sample is mounted in a dilution refrigerator with temperature-controllable base plate in zero magnetic field \cite{Yamahata3}. 

In Fig.~1(c) we show an example trace of $I_{p}$, which shows plateaus for successive $n$ ($n=0$ to 6 are plotted) at temperature $T = 7$~K.
In red, we plot a fit of $I_{p}$ to the decay cascade model \cite{Slava, Kaestner}.
This is an analytic solution to the master equation of the back tunnelling rate, and a fit is given by 
\begin{equation}
  I_{p} = \sum_{n}\exp\left( -\exp\left[ -\delta_{n}\frac{V_{exit} - V_{n}}{V_{n+1}-V_{n}}\right] \right),
\end{equation}
 where $V_{n}$ is the onset in $V_{exit}$ for plateau of population $n$, and $\delta_{n}$ a measure of slope, and hence accuracy.
Because of the appropriateness of this fit, we expect the cross-coupling between $\rm{G_{1}}$ and the QD to be large \cite{Yamahata3}, so that the QD is continuously out of equilibrium.
This is an important criterion for evaluation of our method, and will produce better accuracy than those devices where the QD is in equilibrium \cite{Fujiwarabook, Yamahata3}.
%We evaluate this cross-coupling later, and note that this is an essential requirement to measure the cross-over temperature $T_{0}$.

Figure~1(d) plots the derivative $dI_{p}/dV_{exit}$ for the data shown in Fig.~1(c).
This distribution is representative of the population for successive $n$.
We see the distributions are clearly asymmetric, which is a key feature of the decay cascade model and strong evidence the quantum dot potential varies in time \cite{Slava, Fujiwara2}.
In red on Fig.~1(d) we plot the derivative of the decay cascade fit, which shows excellent agreement.
We would expect non-parabolicity and shape effects to quickly cause the measured distributions of Fig.~1(d) to diverge from the fit, because the analytic solution to the decay cascade model uses a parabolic barrier in its solution of the WKB method in deriving the tunnelling rates \cite{Slava}.
The excellent agreement to data of the decay cascade fit therefore allows us to clearly state that the QD potential profile, and by extension, the form of the entrance barrier created by $\rm{G_{1}}$, is parabolic \cite{Fujiwarabook}.
This claim is strengthened by the accuracy of the fit at high $n$, implying a strong parabolicity to at least $n = 6$.

$\delta_{n}$ characterises the energy dependence of the escape rate across $\rm{G_{1}}$.
It determines the slope of the riser between plateaus (see Fig.~1(c)), and hence is proportional to the flatness of the plateau, making it a valuable measure of a sample's accuracy \cite{Slava,Fujiwarabook,GiblinReview,Yamahata}.
The total error rate of the pump is determined by $\delta_{n}$. 
% and the charging energy $E_{c} \propto V_{n+1}-V_{n}$.
In the following discussion, we will examine $\delta_{1}$, the dominant source of error in pumping a single electron, to deduce the error and evaluate the device potential profile.
To do this, we exploit the temperature dependence of the system.
Figure~2(a) plots $I_{p}$ as a function of $V_{exit}$, equivalent to Fig.~1(c), for successive temperatures $T = 10 - 80$~K for $n = 1$.
The definition of the plateau quickly declines, implying a higher error rate with increasing $T$.
Figure~2(b) plots $\delta_{1}$ for the range of $T$ plotted in Fig.~2(a).
We identify two regimes, the tunnelling limit, at $T < T_{0}$, where the dominant escape mechanism to induce an error is tunnelling across the potential barrier $\rm{G_{1}}$, and the thermal limit $T > T_{0}$, where the dominant mechanism of escape from the dQD is thermally activated barrier hopping \cite{Grabert, Caldeira, Wolynes}.
We define the cross-over temperature $T_{0}$ as the value of $T$ where these rates are equal.
We can express $\delta_{1}$ as $\delta_{1} = \left(1+\frac{1}{g}\right)\frac{E_{c}}{kT^{*}}$, where $g = \alpha_{ent-QD}/\left(\alpha_{ent}-\alpha_{ent-QD}\right)$ is a measure of the cross-coupling between the dQD and $V_{ent}$ ($\alpha_{y} = eC_{y}/C_{\Sigma}$, where $C_{y}$ is the capacitance from source of electric field $y$ and $C_{\Sigma}$ is the total capacitance to the dQD, see Fig.~3(c))\cite{Fujiwarabook}, $k$ is Boltzmann's constant, and $T^{*}$ is an effective temperature \cite{Fujiwarabook,Yamahata3}.
This cross-coupling is sometimes expressed equivalently as the plunger to barrier ratio $\Delta_{ptb} = gkT_{0}$ \cite{Kaestner2}.
Then, $T^{*} = T_{0}$, $T < T_{0}$, and $T^{*} = T$, $T > T_{0}$ and hence the pump's best accuracy (when $T < T_{0}$)  is determined by the tunnel barrier profile, and the charging energy $E_{c}$.
%We expect $\delta_{n} \propto \alpha_{ent}/kT^{*}$, where $\alpha_{ent}$ is the lever arm of gate $\rm{G_{1}}$ on the channel, $k$ is Boltzmann's constant, and $T^{*}$ is an effective temperature \cite{Fujiwarabook,Yamahata3} for any $n$.
%Then, $T^{*} = T_{0}, T < T_{0}$, and $T^{*} = T, T > T_{0}$ and hence the pump's best accuracy (when $T < T_{0}$)  is determined by the tunnel barrier profile, and the charging energy $E_{c}$.
We plot this dependency on Fig.~2(b) as the shaded grey line, finding good agreement for $T_{0} = 17$~K.
%In the low temperature limit ($T < T_{0}$), we can express the tunnelling limited value of $\delta$ as $\delta = \left(1+\frac{1}{g}\right)\frac{E_{c}}{kT_{0}}$, where $g = \alpha_{ent-QD}/\left(\alpha_{ent}-\alpha_{ent-QD}\right)$ is a measure of the cross-coupling between the dQD and $V_{ent}$ \cite{Fujiwarabook}.
The scaled charging energy $\left(1+\frac{1}{g}\right)E_{c}$ is a free parameter in this fit, and we find $\left(1+\frac{1}{g}\right)E_{c} \sim 17.8$~meV, which although an overestimate of $E_{c}$, is in agreement with previous estimates \cite{Fujiwara2,Yamahata4,Yamahata, Yamahata5}.
This suggests $g$ is large, in agreement with the asymmetry of the population profile of Fig.~1(d) \cite{Yamahata3}.
It is very difficult to directly evaluate $g$, because there is no direct mapping between its value and the shape of the distribution.
However, we note that $g = 0$ is a fully static quantum dot (thermal equilibrium condition) and increasing $g$ couples the QD to $V_{ent}$, creating the dynamic quantum dot.
%As $g \rightarrow \infty$, the asymmetry of the distribution (characterised by the ratio of peak heights $\left|d^{2}I_{p}/dV_{exit}^{2}\right|$) tends to $\sim 1.92$ \cite{Yamahata3}.
%In our device, this ratio is (for $T < T_{0}$) $\sim 1.84$.
%In the thermal (unit $g$) case the population distribution is symmetric. 
Because of the asymmetry of the distributions in Fig.~1(d), we can conclude the QD is dynamic and so $g >> 1$. 
We assume $g \sim 10$ \cite{Yamahata3}, and so $E_{c} \sim 15.9 $~meV. 
%In this case, the contribution of $g$ to the error on the evaluation of the charging energy $E_{c}$ is  $\sim 12 $\%.
%The fact that $g$ is not strong can be seen in the cross-over region around $T = T_{0}$ being smooth, contrary to the fit, further supporting our estimation of a mid-ranged value.
%We also note that there can be some contribution throughout due to thermal loading of electrons, akin to a passive QD, which can also reduce the definition of $T_{0}$ and the quality of the decay cascade fit \cite{Yamahata4}.

\begin{figure}
	%\includegraphics[width=\linewidth, scale=0.75]{fig2.png}
%\subfigure
		\includegraphics[width=\linewidth ]{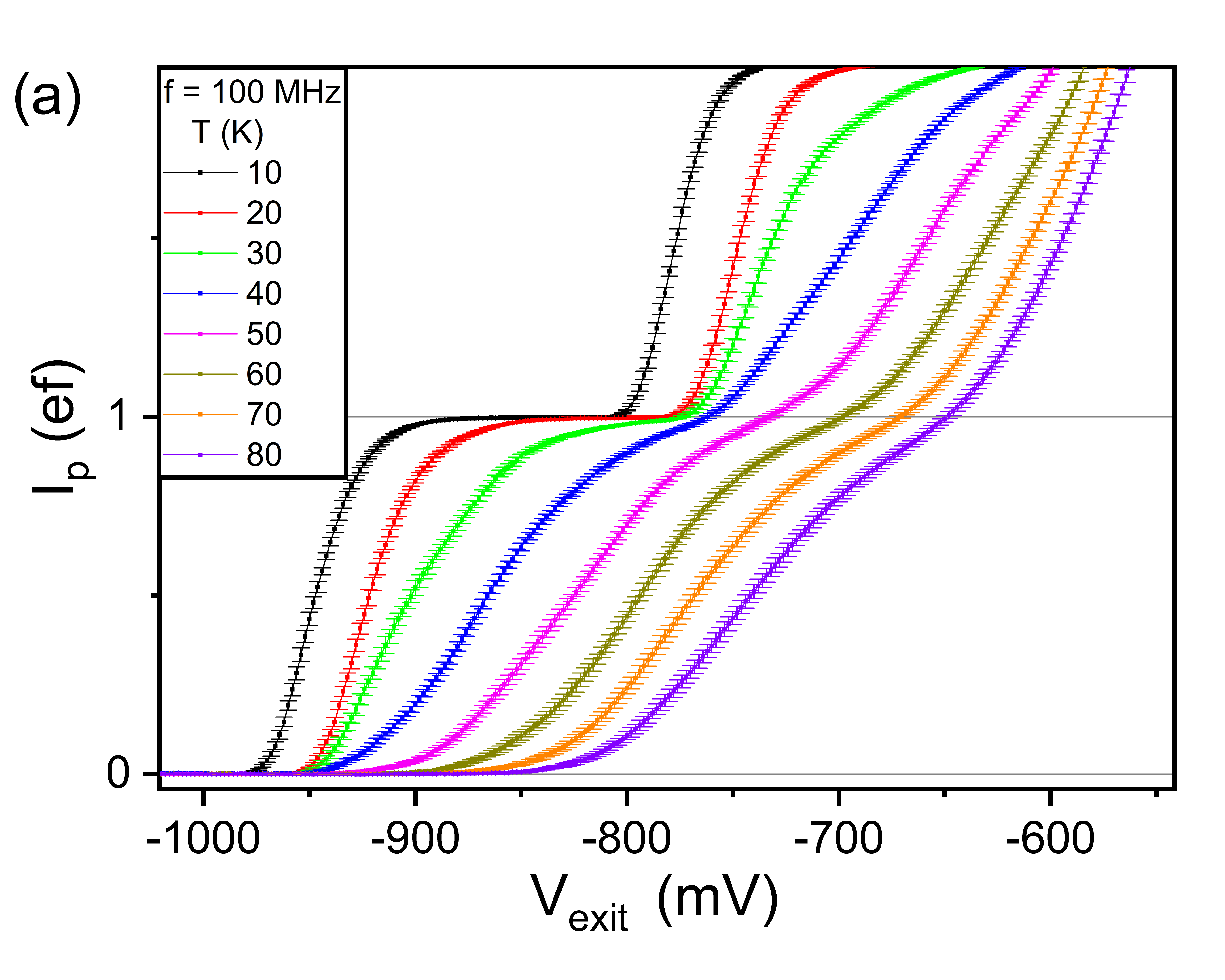}
%\subfigure
	\includegraphics[width=\linewidth ]{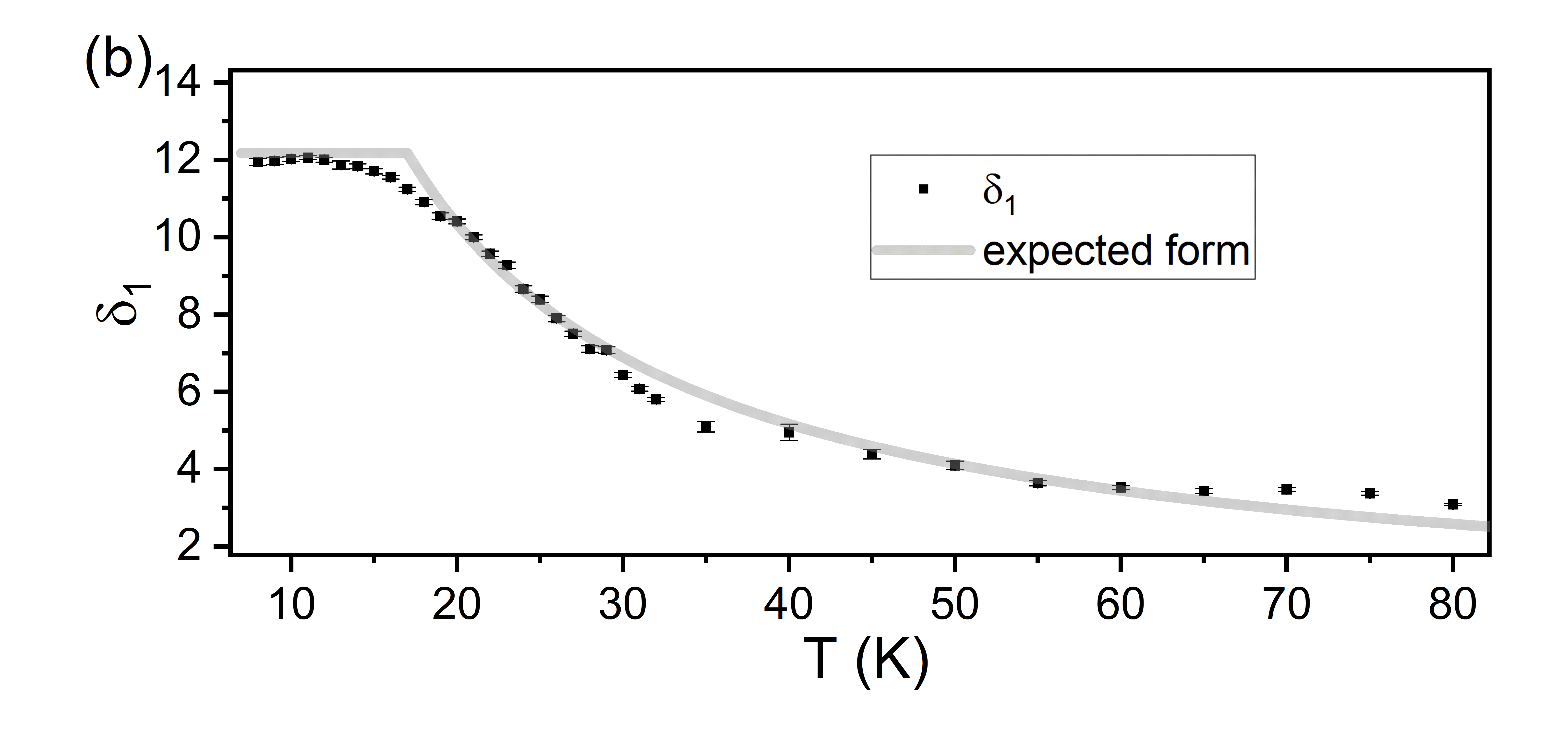}
	\caption{
		(a) $I_{p}$ as a function of temperature $T$. Traces are equivalent to those of Fig.~1(c).
		(b) $\delta_{1}$ as extracted from a fit to the decay cascade model (see Fig.~1(c)) of the traces in (a). In grey, we mark the approximate expected relationship between $\delta_{1}$ and $T$, finding $T_{0} = 17$~K and $E_{c} \sim 15.9$~meV.
	}
	\label{Fig.2}
\end{figure}

From $\delta_{n}$, we can evaluate the lower bound of the error of the loading process, and with $T_{0}$, we can characterise $E_{c}$.
These are two critical values of merit in determining the usefulness of a single-electron wavepacket source.
To begin to have control over these parameters, we need to know the form of the potential profile in the dQD and the barrier formed by $\rm{G_{1}}$.
We can construct this profile from information about the temperature dependence, and $T_{0}$.
To do this, we examine $I_{p}$ at fixed $V_{exit}$, as plotted in Fig.~3(a).
Here, we plot $I_{p}$ as a function of $V_{exit}$ for successive temperatures $T$ from 20 - 80 K, similarly to Fig.~2(a).
We examine the current at the vertical cuts as marked in Fig.~3(a).

In the limit of the validity of the decay cascade model, we can equivalently express $I_{p}$ as \cite{Fujiwarabook,GiblinReview,Kaestner2,Fujiwara2}
\begin{equation}
I_{p} = \exp\left(-\exp\left(-\frac{\alpha V_{exit}}{kT^{*}} + \ln\left(\frac{\Gamma_{1}}{\Gamma_{inc}}\right)\right)\right),
\end{equation}
where $\alpha = \alpha_{exit}\alpha_{ent}/\alpha_{ent-QD}$ is the relative capacitative coupling of the dQD from $V_{exit}$. 
$\Gamma_{1}$ is the total backtunnelling rate when the potential of the dQD crosses the Fermi level $E_{F}$ of the source lead for $V_{exit} = 0$, and $\Gamma_{inc}$ is a measure of the relative change in escape due to the rising of the voltage $V_{ent}^{AC}$.
For simplicity, we will assume $V_{ent}^{AC}$ rises linearly, and assume that $E_{c}$ is sufficiently large that over the temperature range studied here we only have a significant contribution from the $n = 1$ state.
Then, eqn. 1 can be compared to eqn. 2 directly, up to an unknown offset in $V_{exit}$ (this prevents us from evaluating the barrier profile directly from $\delta_{n}$).
This offset arises because the energy of the dQD at the time of formation is unknown.

In Fig.~3(b) we compare the exponent terms by plotting $\ln\left(-\ln\left(I_{p}\right)\right)$ with $T$
\footnote{We note there was a discontinuity seen around $T = 32$~K, arising from a thermal cycle occurring to the device. On remeasuring the device, it was found $\delta_{1}$ was unchanged (ensuring consistency and generality of results) but the position in $V_{exit}$ had slightly shifted by 14~mV. This has been accounted for in this analysis by offsetting curves for $T < 32$~K. Note that the curves of Figs.~2(a) and 3(a) are also offset.}.
%At low $T$, the value of  $\ln\left(-\ln\left(I_{p}\right)\right)$ rapidly changes.
%It then follows a smoother curvature from $\sim 15$~K before changing to a curved dependence at high $T$.
%This can be understood as the combination of two effects.
%Firstly, we note the low $T$ data is adversely affected by the changing conductance due to accumulation at the device back gate interface, changing the nanowire channel conductance \cite{Fujiwara3} (see also similarity of traces for 10 and 20 K in Fig.~3(a)).
%The second effect is the change from a linear shift with $T$ at low $T$ to a non-linear dependence at high $T$.
We perform the fit to the data of Fig.~3(b)
\footnote{In the analysis of Fig.~3(b), low temperature data was excluded from the fit owing to instability of the curve position in $V_{exit}$, which we speculate is due to a changing channel conductance \cite{Fujiwara3}.}
as $
\ln\left(-\ln\left(I_{p}\right)\right) = \ln\left(\Gamma_{V}/\Gamma_{inc}\right) $ with $\Gamma_{V} = \Gamma_{a} \left(\exp\left(-\frac{\Delta}{kT}\right)+\exp\left(-\frac{\Delta}{kT_{0}}\right)/\left(1+\exp\left(-\frac{\Delta}{kT_{0}}\right)\right)\right)$,
and $\Gamma_{inc} = \left(\alpha_{ent}-\alpha_{ent-QD}\right) dV_{ent}^{AC}/dt/\left(kT^{*}\right)$.
Here, $\Gamma_{a}$ is the attempt frequency, $\Delta $ is the entrance barrier height at the point when the dQD level crosses $E_{F}$ (see Fig.~3(c)), and we express $V_{ent}^{AC} = 2\alpha_{ent}ftV_{pp}$ with $V_{pp}$ the peak to peak voltage of $V_{ent}^{AC}$.
%Then $\Gamma_{1} = \Gamma_{V}$ at $V_{exit} = 0$.
We note the most likely time for backtunnelling to occur is just after the dQD potential crosses the lead level, and hence $\Delta$ is a critical parameter in determining the backtunnelling rate.
We define the successive terms in brackets for $\Gamma_{V}$ as a thermal hopping contribution $\left(\exp\left(-\frac{\Delta}{kT}\right)\right)$ and a tunnelling contribution $\left(\exp\left(-\frac{\Delta}{kT_{0}}\right)/\left(1+\exp\left(-\frac{\Delta}{kT_{0}}\right)\right)\right)$ \cite{Yamahata3, inprep}.

\begin{figure}
	\includegraphics[width=\linewidth, scale=0.75]{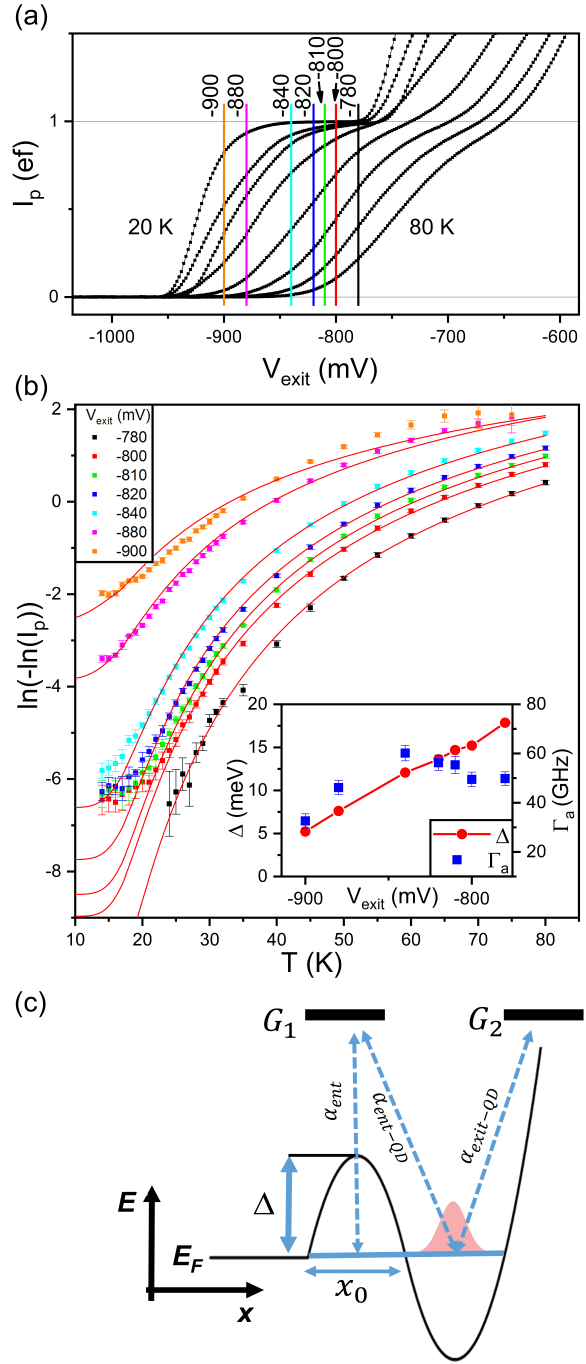}
%\subfigure
	%%	\includegraphics[scale=0.3]{fig3a.png}
%\subfigure
	%	\includegraphics[scale=0.3]{fig3b.png}
%\subfigure
	%	\includegraphics[scale=0.3]{fig3c.png}
	\caption{
		(a) $I_{p}$ as a function of $V_{exit}$ for $T = 20 - 80$ K in 10 K increments (similar to Fig.~2(a)). Vertical lines indicate measurement values of $V_{exit}$ used in the analysis of $\Delta$ in (b). 
		(b) Evaluation of $\Delta$ by plotting  $\ln\left(-\ln\left(I_{p}\right)\right)$ with $T$ for the values of $V_{exit}$ indicated by the colour matched lines in (a), with a fit in red (see main text). Inset: $\Delta$ (left axis) and $\Gamma_{a}$ (right) evaluated for each $V_{exit}$ used in the main plot.
		(c) Reconstruction of the potential profile. $x$ is a real spatial dimension along the length of the channel.
		The $\alpha$ factors are also shown.
	}
	\label{Fig.3}
\end{figure}
%Because of the contribution to the data of Fig.~3(b) from the changing channel conductance, we fit the higher temperature data only, as shown in red on Fig.~3(b).
The fit contains two free parameters, $\Delta$ and $\Gamma_{a}$, which are plotted in the inset to Fig.~3(b).
We expect that $\Delta$ would vary with $V_{exit}$ linearly because of the contribution of the cross capacitance via $\alpha_{exit-QD}$, and we find this to be the case.
$\Gamma_{a}$ should be more constant over this range, because it is dependent more on the shape of the dQD potential profile, and again we find fair agreement with $\Gamma_{a}\sim 50$~GHz.
In the inset to Fig.~3(b), the error plotted is simply the error on the fit, which incorporates the error in the measurement. 
%The scatter in the points likely arises from slight deviation from the expressions made for the thermal hopping and tunnelling approximations made here. 
We note that for $V_{exit} = -780$~mV, which lies close to the region where $n = 2$ dynamics are dominant, the deviation of $\Delta$ from a straight line is minor, as we expect the contribution from $n = 2$ to be at least one order of magnitude smaller even in this close proximity.
In the fit, we have used the value of $T_{0}$ extracted from Fig.~2(b).
%With this value, $V_{exit} = -840, -880$ and $-900$~mV show the crossover from the tunnelling to thermal hopping regime at approximately the same value as the data points.
%For the other values, we note that the data crossover region is not so well fit, implying a different value of $T_{0}$. 
%This is likely because of error in the expression of $\Gamma_{1}$ and instability in $V_{exit}$.
%Hence, because of this apparent confliction in the data, estimation of $T_{0}$ from $\delta_{1}$ is a more accurate method.
We expect that at low temperature, the cross-over region is not well fit due to the measurement uncertainty; $\ln\left(-\ln\left(I_{p}\right)\right) = -6$ corresponds to $I_{p}/ef \sim 0.998$, which is at the limit of our experimental resolution
\footnote{The error bars plotted in Fig.~3(b) are the Type A (statistical) uncertainty, and these are used in the fit. The limit of our measurement resolution, and hence Type A uncertainty, is approximately 0.2~pA, which corresponds to $\ln\left(-\ln\left(I_{p}\right)\right) \sim 6$, and points below this are not used in the fit. To show the crossover to the tunnelling regime, we extend the fit curve in to this region.}.

In Fig.~3(c) we combine our results and sketch the form of the potential profile. 
On the plateau, where the sample should be tuned to find its best possible accuracy for $n = 1$, we find $\Delta \lesssim E_{c}$, as expected before we see the dominance of the $n = 2$ contribution that can be stably permitted for $\Delta = E_{c}$ (see Fig.~1(c)).
We can evaluate the barrier size $x_{0}$ for a given $\Delta$ by expressing the potential due to $V_{ent}$ as $V_{ent} = \Delta - m\omega^{2}x^{2}/2$, with $x$ the spatial dimension along the length of the channel.
$\omega $ can be found from noting that for a parabolic potential $T_{0} = \hbar \omega/2k\pi$ \cite{Yamahata3}.
For our sample we derive $x_{0}= 26$~nm at $V_{exit} = -780$~mV, i.e. at the point of best accuracy.
This is consistent with an expected lithographic length of $\rm{G_{1}}$ of $\sim 40$~nm.

In conclusion, we have presented a method that derives the barrier and dQD potential profile by measuring the backtunnelled current from the dQD at various temperatures.
We can state our method in three key steps: (i) measure the transferred current $I_{p}$ as a function of $V_{exit}$ and fit the decay cascade model (eqn. 1) to establish parabolicity; (ii) extract the parameter $\delta_{n}$ as a function of temperature from the fit to evaluate absolute accuracy and identify the cross-over temperature $T_{0}$; and (iii) fit the measured current as a function of $T$ at fixed $V_{exit}$ to extract $\Delta$, $x_{0}$ and reconstruct the barrier and dQD profile shapes.
 
For engineering the best possible accuracy of the dQD system, $T_{0}$ should be minimised and $E_{c}$ should be maximised, which can be achieved by increasing the gate length of $\rm{G_{1}}$.
Further, we suggest using a non-heterogeneous depth profile of the gate to achieve non-parabolicity in the potential shape. 
\\
\newline
This work was partly supported by JSPS KAKENHI Grant Number JP18H05258.

\bibliography{phonons_reduced}

\end{document}